\documentclass[12pt]{iopart}
\usepackage{color}
\usepackage{graphicx}

\DeclareRobustCommand\openone{\leavevmode\hbox{\small1\normalsize\kern-.33em1}}

\def\bra#1{\mathinner{\langle{#1}|}}
\def\ket#1{\mathinner{|{#1}\rangle}}

\def\x{X}

\def\z{Z}

\def\mx{\mu^x}
\def\my{\mu^y}
\def\mz{\mu^z}

\begin{document}
\title{Generating Topological Order From a 2D Cluster State using a Duality Mapping}
\date{\today}
\author{Benjamin J. Brown}
\address{Institute for Mathematical Sciences, Imperial College London, London SW7 2BW, UK}
\address{Quantum Optics and Laser Science, Blackett Laboratory, Imperial College London, Prince Consort Road, SW7 2AZ, UK}
\ead{benjamin.brown09@imperial.ac.uk}
\author{Wonmin Son}
\address{Center for Quantum Technology, National University of Singapore, 117542 Singapore, Singapore}
\address{Department of physics, Sogang University, Sinsu-dong, Mapo-gu, Seoul 121-742, Korea}
\ead{sonwm@physics.org}
\author{Christina V. Kraus}
\address{Max-Planck-Institute for Quantum Optics, Hans-Kopfermann-Str. 1, D-85748 Garching, Germany}
\author{Rosario Fazio}
\address{NEST, Scuola Normale Superiore, Istituto di Nanoscienze - CNR, Piazza dei Cavalieri 7, I-56126 Pisa, Italy}
\address{Center for Quantum Technology, National University of Singapore, 117542 Singapore, Singapore}

\author{Vlatko Vedral}
\address{Center for Quantum Technology, National University of Singapore, 117542 Singapore, Singapore}
\address{Department of Physics, National University of Singapore, 2 Science Drive 3, Singapore 117542}
\address{University of Oxford, Parks Road, Oxford OX1 3PU, United Kingdom}
\address{Department of Atomic and Laser Physics, Clarendon Laboratory, University of Oxford, Parks Road, Oxford OX1 3PU, UK}

\begin{abstract}
In this paper we prove, extend and review possible mappings between
the two-dimensional Cluster state, Wen's model, the two-dimensional Ising chain and
Kitaev's toric code model. We introduce a two-dimensional duality transformation
to map the two-dimensional lattice cluster state into the topologically-ordered Wen
model. Then, we subsequently investigates how this mapping could be achieved physically,
which allows us to discuss the rate at which a topologically ordered system can
be achieved. Next, using a lattice fermionization method, Wen's model is mapped
into a series of one-dimensional Ising interactions. Considering the boundary terms
with this mapping then reveals how the Ising chains interact with one another.
The relationships discussed in this paper allow us to consider these models from
two different perspectives: From the perspective of condensed matter physics these
mappings allow us to learn more about the relation between the ground state properties
of the four different models, such as their entanglement or topological structure.
On the other hand, we take the duality of these models as a starting point to address
questions related to the universality of their ground states for quantum computation.
\end{abstract}

\maketitle

\section{Introduction}
Studying the physical properties of lattice spin systems is one of the most challenging, yet most fascinating areas of condensed matter physics. These systems exhibit a plethora of exotic quantum phases, some of which are even believed to be useful for quantum computation. Those proposals have in common that they exploit special ground state properties of a spin system, e.g. its entanglement or topological structure, to overcome the notorious problems of the standard circuit model of quantum computation, such as the individual control of qubits or decoherence. Two prominent examples of such proposals are  Measurement-based Quantum Computation (MBQC) and Topological Quantum Computation (TQC).

In the case of MBQC, the computational task is achieved via single qubit measurements on a highly-entangled state \cite{Briegel09}. These measurements are not only much simpler to realize experimentally, but this approach also reduces the problem of quantum control, because it is not necessary to interact qubits after the state is prepared. If the entangled state is appropriately chosen, MBQC is equivalent to the standard circuit model, i.e. it provides us with a universal quantum computer. One example of such a state in two dimensions  is the so-called Cluster State \cite{Raussendorf01}.

TQC, on the other hand, uses a two-dimensional topologically ordered system with non-Abelian braiding statistics, and braids the quasi-particle excitations \cite{Sarma05} in such a way that logical qubits are processed to perform computational tasks \cite{Nayak}. This model of quantum computation is robust against errors because the long-range entanglement of the system prevents perturbations of small areas of physical qubits from causing errors to the logical qubit. A simple model with topological order and Abelian braiding statistics which may help us better understand TQC is Kitaev's toric code model \cite{Kitaev03, Dennis}, which was originally designed to obtain a topologically protected quantum memory.

\begin{figure}[htbp]
\begin{center}
\includegraphics[width=4.5in]{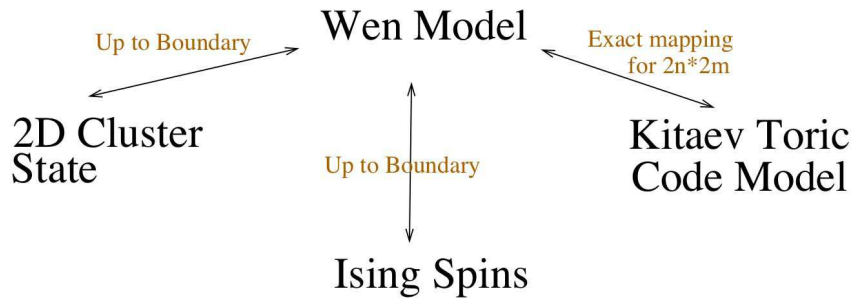}
\end{center}
\label{fig:Mapping}
\caption{Summary of the mapping between the two-dimensional cluster state, the Wen model, the Ising model and Kitaev's toric code model. The two-dimensional cluster state is mapped up to the boundary terms to Wen's model using a duality transformation. Wen's model can be diagonialized using a fermionization approach leading, up to boundary terms, to a Hamiltonian of anti-ferromagnetic Ising spins. On the other hand, the Wen model can be mapped exactly into Kitaev's toric code model when the lattice is of even by even dimensions.}
\end{figure}

While these two proposals make use of two very different properties of the systems to achieve quantum computation, they share the rare property that they are the ground states of some non-trivially interacting exactly-solvable spin system. The present work shows a method of directly mapping between the two-dimensional cluster state on a square lattice and Wen's model, which in turn relates the cluster state with the toric code \cite{Nussinova09} and the Ising model \cite{Chen07}. In that case, the boundary terms transform nontrivially and the ground state degeneracy is changed. These mappings are summarized in figure \ref{fig:Mapping}. This work can be considered a generalisation of the one-dimensional work done in reference \cite{Son}, which transforms the one-dimensional Cluster State into the one-dimensional Ising model. This particular work considers the boundary terms and topological properties of these models.

Motivated by the importance of two-dimensional spin models for the realization of a quantum computing device, and fascinated by the possibility to map these models which seem unrelated at a first glance onto each other, the goal of this work is to review and extend known results as well as to prove new possible mappings between these systems using our new mapping between the cluster state and Wen's model, and some existing mappings.

This paper is organized as follows:  We start with a review of the cluster state model and study its topological properties. Next, we prove that in the thermodynamic limit the two-dimensional cluster state model can be mapped into the topologically ordered Wen's model \cite{Wenmodel} using a global unitary transformation. In the case of open boundary conditions, Chen and Hu \cite{Chen07} have shown how Wen's model can be diagonalized using a Jordan-Wigner transformation \cite{Jordan28, Katsura61}. We first review their technique and then turn our attention to the more involved case of periodic boundary conditions. We do not only give a constructive way how to obtain the spectrum of the Wen Hamiltonian in this way, but we also include a discussion about the ground state degeneracy. Finally,  we review the local unitary transformation that takes Wen's model into Kitaev's toric code model \cite{Nussinova09}, establishing a link between the two-dimensional cluster state and Kitaev's toric code. This relationship is then used to interpret the boundary terms of the mapping between the cluster state and Wen's model when the cluster state on a square periodic lattice is transformed.

\section{The Family of Cluster States}

Consider a $D$-dimensional square lattice and associate to each site a spin-half particle. Then, the cluster state, also known as a graph state when it's on a square lattice, on that lattice is defined as the common eigenstate of the ``stabilizers" $S_i= X_{i}
\prod_{j \in {\cal N}(i)}Z_{j}$ with eigenvalue $+1$
$\forall i$. Here, $X_{i}$ and $Z_i$ are the Pauli
spin operators acting on the $i$-th site and ${\cal N}(i)$ denotes the neighboring sites of the
$i$-th spin. Any cluster state is the non-degenerate ground state of the Hamiltonian $H_C=-\sum_i S_i$ with eigenvalue $-N$.

More insight into the structure of the family of cluster states can be obtained using its operational construction: A cluster state can prepared via an application of controlled phase gates $U_{i,j}=[\openone
+Z_i+Z_{j}-Z_i Z_{j}]/2$ to the product state $|+\rangle^{\otimes N}$ in the following way:
\begin{equation}\label{eq:def_cluster}
|C\rangle=\left(\prod_{i} U_{i,{\cal N}(i)}\right)
|+\rangle^{\otimes N}.
\end{equation}

The operational representation allows us to write down immediately the complete spectrum of $H_c$: If we define $|C_{\vec{n}}\rangle =
\prod_{j}(Z_j)^{n_j}|C\rangle$, where
$\vec{n}\equiv(n_1,n_2, \cdots,n_N)$ and $n_j\in\{0,1\}$, then one can easily show that $S_i |C_{\vec{n}}\rangle=(-1)^{n_i}|C_{\vec{n}}\rangle$ $\forall i$ and that $\langle
C_{\vec{n}}|C_{\vec{m}}\rangle=\delta _{\vec{n},\vec{m}}$ holds. Further, we can read off that the energy eigenvalue of $|C_{\vec{n}}\rangle$  is a function of the number of spin flips,  $E_{\vec n}=-N+2\sum_i n_i$.

The cluster state is a highly entangled state in a sense that the state has the largest relative entropy  that any deterministic state can reach for a given number of qubits \cite{Damian07}. The unit of cluster state entanglement is $N/2$ in terms of the relative entropy of entanglement \cite{Vedral02} and it scales like $N$ in other measures of multipartite entanglement
as well \cite{Jungnitsch10}.

Next, we discuss the topological properties of the family of cluster states defined in Eq.~(\ref{eq:def_cluster}).
As it has been proposed in reference~\cite{Levin06} and \cite{Kitaev06}, the von Neuman entropy serves as a measure
to characterize the topological order of a many-body system $S$ in the following way: Let $A$ be a subsystem
of $S$, and let $L =|\partial A|$ be the length of the boundary between $A$ and $S\setminus A$.
Then, the von-Neuman entropy is given by
\begin{equation}\label{eq:def_top_entropy}
S(|\psi\rangle)= - \mbox{Tr}(\rho_A \log \rho_A) = \alpha L-\gamma+\cdot\cdot\cdot,
\end{equation}
where $\rho_A$ denotes the reduced density matrix of the subsystem $A$. The ellipsis represent terms that vanish in the limit of $L\rightarrow \infty$. The scale invariant part $\gamma$ characterizes the global feature of the entanglement in the ground state, called \emph{topological entanglement entropy}.

Making use of the operational representation of the cluster state given in Eq. (\ref{eq:def_cluster}), the topological entanglement entropy of the cluster state can be readily calculated. To this end, we make use of the fact that any cluster state can be obtained via a local application of controlled phase-gates $U_{k}^{cp}$ and that every individual spin of the system is in a completely mixed state. If we further denote as $\cal N$ those neighboring sites of the boundary $\partial A$ which lie outside of region $A$ we have that
$\rho_A=\mbox{Tr}_{\cal N}\Big[ {\bf U}\openone_{\cal N}\otimes |C\rangle_A\langle C|{\bf U}^{\dagger}\Big]$.
Here, $|C\rangle_A$ denotes the cluster state defined on subsystem $A$ alone, $\openone_{\cal N}$ is the state of the spins in boundary $\cal N$ and ${\bf U}=\left(\prod_{k\in \partial A }U_{k}^{cp}\right)$ is the unitary of the controlled phase gates applied to the spins on $\partial A$ and $\cal N$. Thus, $
\rho_I=\frac{1}{2^{L}}\sum_{\{n_k\in \partial A\}=0}^1  \hat{B}_{\bar{n}}|C\rangle_I\langle C|\hat{B}_{\bar{n}}$,
where $\hat{B}_{\bar{n}}=\prod_{k\in \partial A}(Z_{k})^{n_k}$, $\bar{n}=(n_1,n_2,\cdots)$, and
$L=\sum_{k\in \partial A}$ is the number of qubits in the boundary $\partial A$. In this case, the boundary $\partial A$ is defined.
With this result at hand, we use Eq.~(\ref{eq:def_top_entropy}) to arrive at
\begin{equation}
S(|C\rangle)= - \frac{1}{2^{L}} \log \frac{1}{2^{L}} \sum_{\{n_k\in ~boundary\}=0}^1 = L.
\end{equation}
A comparison with Eq.~(\ref{eq:def_top_entropy}) reveals that the topological entanglement entropy $\gamma = 0$.

Thus, a cluster state on an arbitrary graph is the unique ground state of a stabilizer Hamiltonian without topological order. However, as we will show in the next section, the two-dimensional cluster state is linked to topologically ordered systems via a simple duality transformation. This seems like a contradiction because the energy spectrum will not be conserved between a cluster-state model with a non-degenerate ground state and a topologically ordered model. This is resolved later in the paper where we consider boundary terms.

\section{Mapping of the Two-Dimensional Cluster State to a Topologically Ordered System}

In this section, we show how the two-dimensional cluster state Hamiltonian can be transformed into a Hamiltonian with topological order using a simple unitary transformation. Recall that the Hamiltonian of the two-dimensional cluster state on a square lattice is of the form
\begin{equation}
H_{C}^{2D}=-\sum X_{i,j} Z_{i-1,j}Z_{i+1,j}Z_{i,j-1}Z_{i,j+1}.
\end{equation}
The duality mapping between the cluster-state model and Wen's model can be written as follows: Firstly, the indices are transformed $(i,j) \rightarrow (i',j')$ such that $j=j'$ and $ i' = i -j +1$. The effect of this index transformation is shown in figure 2.
\begin{figure*}
\begin{center}
\begin{tabular}{ccc}  {\bf a.} & $ \qquad $& {\bf b.}
\\
\includegraphics[scale=2.5]{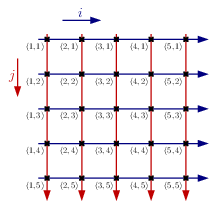} &  &
\includegraphics[scale=2.5]{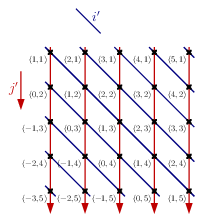}
\end{tabular}
\end{center}
\caption{{\bf a.} Shows the standard coordinates, the arrowheads show the direction in which the coordinates increase. {\bf b.} Shows the new coordinates on a $6 \times 6$ lattice. The red arrows shows the $j'$ coordinates still increase, but the diagonal blue lines don't have arrowheads because they show lines under which the $i'$ coordinates remain constant. \label{coord}}
\end{figure*}
Then the mapping can be conveniently written
$$ \mx_{i',j'} = \z_{i',j'} \z_{i',j'+1}, $$
and
$$ \mz_{i',j'} = \prod_{k=1}^{j'} \x_{i',k}, $$
where the modes are ordered in a diagonal way on the lattice as depicted in Fig \ref{fig:Diag}.
\begin{figure}[tbp]
\begin{center}
\includegraphics[width=2in]{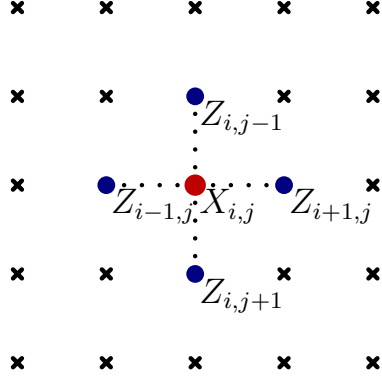}
\end{center}
\label{fig:Diag}
\caption{Graphical representation of one term $X_{i,j} Z_{i-1,j}Z_{i+1,j}Z_{i,j-1}Z_{i,j+1}$ of the 2D Cluster Hamiltonian.}
\end{figure}

A straightforward calculation shows that under this dual transformation the cluster state Hamiltonian is transformed into Wen's model \cite{Wenmodel}
\begin{equation}
H_{W}=-\sum \mu_{i,j-1}^z \mu_{i-1,j}^z \mu_{i-1,j-1}^x\mu_{i,j}^x,
\end{equation}
where we have now reverted back to the original indices.

Wen's model is of special interest since it exhibits a new kind of phase transition where a change of the sign in the Hamiltonian leads to a change of the quantum order of the system while the symmetry is preserved. The model is different from Kitaev's toric code model \cite{Kitaev03}, but it can be mapped into it  through a local unitary transformation. Further, Wen's model corresponds to the low energy sector of Kitaev's honeycomb lattice model in the limit $J_z \gg J_x, J_y$ \cite{Wenmodel}.

This mapping can be discussed in more depth by considering the physical operation that performs such a transformation. The mapping is done using the following circuit \cite{Plenio} along each of the diagonals of a periodic lattice of $N\times N$ spins.
\begin{eqnarray*}U_{i'}   &=&\left[ \prod_{j=1}^{N-1}\left( \openone_{i',N-j } \otimes \ket{0}\bra{0}_{i', N-j+1} + \x_{i',N-j } \otimes \ket{1}\bra{1}_{i', N-j+1}    \right) \right]^T   \\ && \times   \left[ \prod_{j=1}^N \frac{1}{\sqrt{2}}\left( \x_{i',j} + \z_{i',j} \right) \right] \end{eqnarray*}
 The transpose acting on the product of the controlled-not gates reverses their order (of course the controlled-not gate is symmetric, so all the transpose does is reverse the order of the individual gates).  The order is important because all the controlled-not gates have to be performed sequentially. They cannot be performed simultaneously because they don't commute with one another on pairs of nearest neighbour sites. With this in mind, and assuming that each of the controlled-not gates requires only one unit of time to be performed, and all the Hadamard gates can be performed simultaneously (as they all occur on different sites) and each of the $U_{i'}$ operators can be performed simultaneously, then the time it takes to transform a cluster state into a topologically ordered system scales linearly with the length of the boundary of the lattice. This agrees with the result found in reference \cite{Bravyi} who find that the optimal time to generate a topologically ordered system scales with the boundary of the system, and not with the volume of the system. This matches the rate that topological order is generated in reference \cite{Dennis}, which is also mentioned in reference \cite{Bravyi}.

\section{Wen's Model and its explicit solution}

Wen's model describes spins on a two-dimensional lattice which are subject to spin-spin interactions on $2\times2$-plaquettes~\cite{Wenmodel},
\begin{equation}\label{eq:H_Wen}
\hat{H}_{\cal W} = g \sum_{i,j} W_{ij},
\end{equation}
where $W_{ij} = X_{ij} Z_{i+1,j} Z_{i,j+1} X_{i+1,j+1} $ and $g$ is the interaction strength. This model was designed to be exactly solvable using a projective construction, which is the same method adopted by Kitaev to solve the honeycomb model \cite{Kitaev06}. This paper uses an alternative approach to diagonalize this Hamiltonian introduced by Chen and Hu in Ref~\cite{Chen07} and generalizes this approach to periodic boundary conditions. This method uses the Jordan-Wigner transformation to write the spin-half operators in terms of fermionic creation and annihilation operators
\begin{equation}
2 c_{ij}^{\dagger}=\left(\prod_{j'< j}\prod_{i'} Y_{i'j'}\right)
\left(\prod_{i'< i} Y_{i'j}\right)(X_{ij}+i Z_{ij}),
\end{equation}
that obey canonical anti commutation relations (CAR) $\{c^{\dagger}_k, c_l\}=\delta_{kl}$.
Equivalently, one can use the hermitian Majorana operators
\begin{equation}
\alpha_{ij}=-i(c_{ij}^{\dagger}-c_{ij}),~~~\beta_{ij}=c_{ij}^{\dagger}+c_{ij},
\end{equation}
that also fulfill the CAR. Now we define new fermionic modes as  linear superpositions of Majorana fermions on two neighboring sites $(i,j)-(i, j+1)$, $d_{ij}=(\alpha_{ij}+i\beta_{i,j+1})/2$. With these new modes, it follows that
\begin{equation}
i \alpha_{ij} \beta_{i,j+1} = 2 d_{ij}^{\dagger} d_{ij}-1 = X_{ij}\Big(Y_{i+1,j} Y_{i+2,j}\cdots Y_{i-1, j+1}\Big)X_{i,j+1},
\end{equation}
and the interaction operators in Wen's model take the simple form
\begin{equation}
W_{ij}=\left(2 d_{ij}^{\dagger} d_{ij}-1\right)\left(2 d_{i+1,j}^{\dagger} d_{i+1,j}-1\right).
\end{equation}

Wen's model transforms differently, depending on the boundary conditions.
We consider first an $N\times M$ lattice with open boundary conditions.
Here, the Hamiltonian Eq.~(\ref{eq:H_Wen}) takes the form
\begin{equation}\label{eq:Hopen}
H_{\cal W}^{\mbox{open}}=g \sum_{j=1}^{M-1}\sum_{i=1}^{N-1} (2\hat{\cal N}_{ij}-1)(2\hat{\cal N}_{i+1j}-1),
\end{equation}
where $\hat{\cal N}_{ij}\equiv d_{ij}^{\dagger} d_{ij}$. In this case, neighboring fermions
in each row are coupled by an Ising-type interactions. However there are no inter-row interactions
so that the Hamiltonian takes the form of $M-1$ independent Ising chains. With this in mind,
it is obvious that each row has a two-fold degenerate ground state
\begin{eqnarray}
\label{eq:groundising}
|\psi^+\rangle_j= \prod_{i=1}^{N/2} d_{2i j}^{\dagger}|\Omega\rangle, ~~~~
|\psi^-\rangle_j= \prod_{i=1}^{N/2} d_{2i-1 j}^{\dagger}|\Omega\rangle,
\end{eqnarray}
where $|\Omega\rangle$ is the state of the vacuum. The states $|\psi^+\rangle_j$ and
$|\psi^-\rangle_j$ have anti-ferromagnetic excitations in the fermionic bases.
The ground state of the Hamiltonian is then given by
\begin{equation}
| \psi_{\bar{n}}\rangle = \bigotimes_{j=1}^{M} |\psi^{n_j}\rangle_j,
\end{equation}
where $\bar{n}=(n_1, n_2, \cdots )$ and $n_j\in\{+,-\}$. \footnote{Note, that if we defined the Hamiltonian Eq. (\ref{eq:Hopen}) for periodic boundary conditions, i.e. on a torus, the ground state would be $2^{M}$-fold degenerate. However, if we interchanged rows and columns, the degeneracy would change to $2^N$, as we now have a torus of a different size.}

In the case of periodic boundary conditions the diagonalized Hamiltonian takes a different form.
The periodicity of the system written in terms of spin-half operators imposes the constraint
$W_{i+N, j+M}=W_{i, j}$. Then, using the same fermionic transformation rules, the Hamiltonian takes the form
\begin{equation}
\label{eq:periodic}
H_{\cal W}^{\mbox{periodic}}=g \sum_{j=1}^{M}\sum_{i=1}^{N-1} (2\hat{\cal N}_{ij}-1)(2\hat{\cal N}_{i+1j}-1)+g \sum_{j=1}^{M} \hat{L}_j,
\end{equation}
where $\hat{L}_j$ is of the form
\begin{equation}
\hat{L}_j=\prod_{i=2}^{N-1}(2\hat{\cal N}_{ij}-1)\prod_{i=1}^{N}\beta_{ij}\prod_{i=1}^{N} \alpha_{ij+1},
\end{equation}
introducing an interaction between row $j-1$ and $j+1$ (because the $\beta_{ij}$ operator affects
the $(i,j-1)$-th site). Since the operators $\alpha_{ij}$ and $\beta_{ij}$ flip the excitation of
the fermion mode $d_{ij}$, it follows that for even $N$
\begin{equation}
\prod_{i=1}^{N}\alpha_{ij}|\psi^{\pm}\rangle_j=|\psi^{\mp}\rangle_j,~~~~~~~~ \prod_{i=1}^{N}\beta_{ij}|\psi^{\pm}\rangle_{j-1}=|\psi^{\mp}\rangle_{j-1}.
\end{equation}
For the sake of simplicity, we deal only with the case that $N$ is even in this paper. Then, the boundary operators impose that we have to superpose the two eigenstates $|\psi^{\pm}\rangle_j$ in each row,  and one of the ground state of the Hamiltonian (\ref{eq:periodic}) is given by
\begin{equation}
|\Psi_0^{(1)}\rangle =\bigotimes_{j=1}^{M} \left(|\psi^{+}\rangle_j \pm |\psi^{-}\rangle_j\right).
\end{equation}
The two ground states can be seen as the product of a Greenberger-Horne-Zeilinger type entangled state \cite{GHZ}. Further, in the case of periodic boundary conditions and an even number of rows (i.e. $M$ even), $|\Psi_0^{(1)}\rangle$ is not the unique ground state of Wen 's model, but we have two other solutions. The complete set of ground state is given by %
\begin{equation}
|\Psi_{0}\rangle=\bigotimes_{j=1}^{M/2} \left(|\psi^{+}\rangle_{2j}\pm|\psi^{-}\rangle_{2j}\right)
\left(|\psi^{+}\rangle_{2j+1}\pm|\psi^{-}\rangle_{2j+1}\right).
\end{equation}
The ground states of Wen's model can be understood using the stabilizer formalism \cite{NandC}. In the case where $g<0$, the ground states of the model will be stabilized by all of the terms in the Hamiltonian $W_{ij}$. In fact, these terms will form a generating set for the entire stabilizer group. However, in the case of infinite or periodic boundary conditions, the operators $W_{ij}$ do not form an independent set because of the condition that $\prod_{i,j} W_{ij} =\openone$. This implies that $\prod_{i\not=k,j\not=l} W_{ij} = W_{kl}$ and if one of the interaction operators is removed, then the generating set will become independent. Following the stabilizer formalism, if a stabilizer group describing $N$ qubits contains $k$ generators in the generating set, then the stabilizer group will describe $2^{N-k}$ states \cite{NandC}. As there are $N$ operators $W_{ij}$, of which $N-1$ are independent, it is clear that in the general case periodic lattices will be two-fold degenerate. In the special case that lattice is either infinite, or the lattice has dimensions of an even by even number of spins then the interaction operators also follow the condition that $\prod_{i+j=\mbox{even}} W_{ij}= \openone$. This means that there are only $N-2$ independent operators in the generating set, so it follows that the ground state will be four-fold degenerate.

\section{Connection to Kitaev's Toric Code Model}

In this section, we demonstrate that Wen's model can be transformed into Kitaev's toric code model using only local unitary transformations. Since local transformations have no influence on topological effects, this allows us to evaluate the entanglement of Wen's model by inspecting the ground state of the toric code model. This equivalence, coupled with the duality mapping also provides the connection between the two-dimensional cluster state and Kitaev's toric code model. The connection between Wen's model and Kitaev's toric code model has been studied by Nussinov and Ortiz \cite{Nussinova09}. We investigate this transformation for periodic boundary conditions, and look for the necessary conditions for the transformation to be faithful and when the mapping is mismatched.  Moreover, we show how to construct the exact ground states of Kitaev's toric code model in a spin basis.

The toric code describes a family of simple spin systems with local interactions
in which the existence of anyons can be demonstrated.
Due to their braiding statistics the state can been used for topological
quantum computation~\cite{Kitaev09}. Kitaev's toric code model on a square lattice,
\begin{equation}
H_{K}=-\sum_{v} A_{v}-\sum_{p} B_{p},
\end{equation}
is a sum of constraint operators associated with vertices $v$ and plaquettes $p$, namely
\begin{equation}
A_v=\prod_{j\in v} X_j, ~~~ B_p=\prod_{j\in p} Z_j,
\end{equation}
where $v =  \{ (i,j),  (i+1,j),  (i,j+1),(i+1,j+1)\}, \quad \forall i,j$ where $ i+j = \mbox{even}$ and $p=\{ (i,j), (i+1,j), (i,j+1), (i+1,j+1) \},\quad \forall i,j$ where $ i+j=\mbox{ odd}$.

\begin{figure}[htbp]
\begin{center}
\includegraphics[width=3.5in]{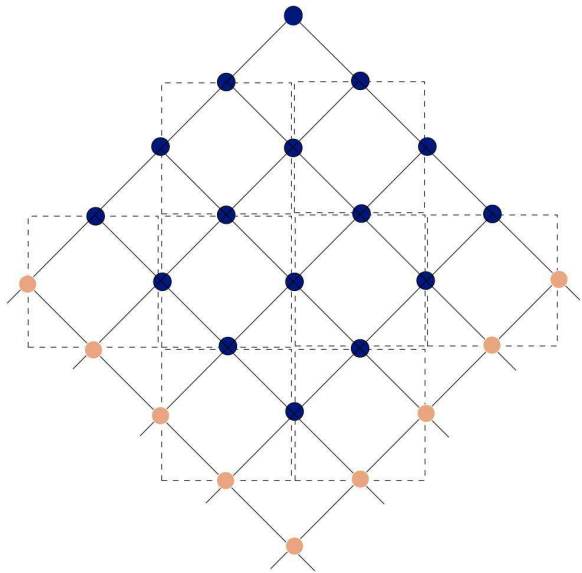}
\end{center}
\label{fig:Trans}
\caption{Transformation of a four-by-four square lattice from Wen's spin model into Kitaev's toric code model. The blue dots are the actual spins in the four-by-four lattice and the yellow dots are the spins which are translated from the first row and the first column. The solid line represents the lattice for the Wen model and the dashed lines denote the transformed lattice in the toric code model. They are both periodic under translation.}
\end{figure}

Wen's model can be mapped into the toric code using a completely local unitary transformation, $U H_{\cal W} U^{\dagger} =H_K$ where
\begin{equation}
U=\prod_{i+j=even} \frac{1}{\sqrt{2}} \left( X_{i,j}+Z_{i,j}\right).
\end{equation}
Note, that for periodic boundary conditions this relation is only valid for lattices with an even number of sites in each direction. Let us briefly explain the action of this unitary: If $i+j$ is even, $UW_{ij}U^{\dagger} = A_{ij}$, while for $i+j$ odd we find $UW_{ij}U^{\dagger} =B_{ij}$ which is a plaquette on $p$. The conventional method of graphically representing the toric code represents the spins as edges on a graph, instead of using the vertices \cite{Kitaev03, Dennis}. One can see that this is equivalent by replacing the vertices with diagonal lines. If we then replace the spins on even sites with diagonal lines that go from the top left corner to the bottom right corner, and spins on odd sites with diagonals that go from the bottom left corner to the top right corner then we have a new graph where edges represent spins. This is represented in figure \ref{fig:Trans} on a 4$\times 4$ lattice. One can see that, to obtain a faithful mapping from Wen's model to the toric code model, Wen's model has to be on a lattice of an even number of sites to an even number of sites. If this is not the case then the toric code picture will not have alternating diagonal lines across the boundary terms, and there will not be perfect star and plaquette operators which consist of only one Pauli operator across the boundaries.

Having these results at hand, we turn our attention now towards the ground state of the toric code. We start with an explicit construction of the ground states of the toric code in the case of periodic boundary conditions. Let $N_S$ denote the number of spins on the lattice. One of the four ground states of the toric code can be constructed using the set of vertex operators $A_v$  ($v=1,\dots,\frac{N_S}{2}-1$) as a generating set\footnote{Note that there are in fact $ \frac{N_S}{2}$ star operators because there are two qubits per vertex. However, due to the boundary condition $\prod_{v}A_v =  \openone$, only $\frac{N_S}{2}-1$ of them are independent.} in the following way:
\begin{equation}\label{eq:gs_toric_code}
\left| \psi_0 \right\rangle = \frac{1}{\sqrt{2^{\frac{N_S}{2}-1}}} \sum_{\bar{n}} \prod_{k=1}^{\frac{N_S}{2}-1} A_k^{n_k} \left| \Omega \right\rangle,
\end{equation}
where $\bar{n} =(n_1,n_2, \dots n_{\frac{N_S}{2}-1})$, $n_i\in \left\{ 0,1 \right\}$ and $ \left|\Omega \right\rangle$ denotes the vacuum in the $Z$-basis.

To check that this is indeed the ground state of the toric code, we have to show that $|\psi_0\rangle$ is a stabilizer state of the sets $\{A_v\}_v$ and $\{B_p\}_p$. We start with the plaquette operators. Using the fact that all vertex and all plaquette operators commute, we find immediately that
 $$ B_p \left| \psi_0 \right\rangle =  \frac{1}{\sqrt{2^{\frac{N_S}{2}-1}}} \sum_{\bar{n}} \prod_{k=1}^{\frac{N_S}{2}-1} A_k^{n_k} B_p \left| \Omega \right\rangle =  \frac{1}{\sqrt{2^{\frac{N_S}{2}-1}}} \sum_{\bar{n}} \prod_{k=1}^{\frac{N_S}{2}-1} A_k^{n_k} \left| \Omega \right\rangle =\left| \psi_0 \right\rangle \quad \forall p .  $$
It is slightly more complicated to show that $| \psi_0\rangle$ is also a stabilizer state for the vertex operator. To this end, we rewrite the ground state for some fixed $v$ in the following way:
$$
\left| \psi_0 \right\rangle = \left(  \openone+A_v \right)  \frac{1}{\sqrt{2^{\frac{N_S}{2}-1}}} \sum_{\bar{n}_v} \prod_{k=1,k\not=v}^{\frac{N_S}{2}-1} A_k^{n_k} \left| \Omega \right\rangle ,
$$
where $ \bar{n}_v = (n_1,n_1,\dots, n_{v-1}, n_{v+1}, \dots, n_{\frac{N_S}{2}-1}) $. Since $A_v^2 = \openone$, we see immediately that $A_v \left| \psi_0 \right\rangle = \left| \psi_0 \right\rangle$ holds.

The other three ground states $\left| \psi_1 \right\rangle $, $\left| \psi_2 \right\rangle $ and $\left| \psi_3 \right\rangle $ can be found by applying non-contractable loops $w_1$ and $w_2$ composed of Pauli $X$ operators around the torus in the horizontal and vertical directions such that the other ground states are $ \left| \psi_1 \right\rangle = w_1 \left| \psi_0 \right\rangle  $, $\left| \psi_2 \right\rangle =w_2\left| \psi_0 \right\rangle $ and $\left| \psi_3 \right\rangle = w_1 w_2 \left| \psi_0 \right\rangle $. The non-contractible loop operators commute with all of the vertex and plaquette operators, thus it is trivial to show that these are the other ground states of the toric code Hamiltonian.

\section{Interpreting the Mapping from the Cluster-State to Wen's model on a Square Periodic Lattice}
In this section, we investigate the duality mapping between the Cluster state model and Wen's model  for periodic boundary conditions. Here, special care has to be taken since the spins are ordered on the lattice in a diagonal way. For simplicity, we only consider the case of an $N \times N$ lattice, where $N$ is even. When considering periodic boundary conditions, we need to be more precise with how we define our indices. We the site $(N+k,j') = (k,j')$ and the site $(1-k,j') = (N+1-k,j') $. Then, to ensure the new spin operators $\mx$, $\my$ and $\mz$ obey the $SU(2)$ commutation relations, the mapping has to be changed as follows; when $j'=N$ is $\mx_{i', N} = \z_{i',N}$ for all $i'$. Otherwise, all the mappings are the same.

Having modified the mapping for square periodic lattices, it is then possible to map the boundary terms for the cluster state. First, it is easy to show that when $  2 \le j \le N-1$, the terms $C_{1,j}$ and $C_{N,j}$ map exactly into Wen plaquettes, $C_{1, j} =  \mz_{N,j-1} \mx_{1,j-1} \mx_{N,j} \mz_{1,j}, $ and $C_{N, j}  =   \mz_{N-1,j-1} \mx_{N,j-1} \mx_{N-1,j} \mz_{N,j}$ using the standard indices.

The other boundary terms are not as easy to interpret. We have that  $$C_{i, 1} =  \mx_{i-1,1} \mz_{i,1} \left[ \prod_{k=1}^{n-i} \mx_{i+k,k} \right] \left[ \prod_{k=1}^{i-1} \mx_{k,n-i+k} \right], $$ when $j=1$ and $$C_{i, N} = \mz_{i-1,N-1}\mx_{i,N-1}  \mz_{i,N} \left[ \prod_{k=0}^{N-i} \mx_{i+k, 1+k}  \right] \left[ \prod_{k=1}^{i-2} \mx_{k,N-i+1+k} \right], $$ when $j=N$.

While mapping these boundary terms lead to quite unintuitive terms in the Hamiltonian, the ground state of this Hamiltonian can interpreted a little easier by manipulating the stablizers. Firstly, one can replace the generators of the stablizer group $ C_{i,N}$ with the stablizers $ C_{i-1,1}C_{i,N}$ $\forall i$. Then in the Wen picture we have that $C_{i-1,1}C_{i,N} =  \mx_{i-2,1} \mz_{i-1,1} \mz_{i-1,N-1} \mx_{i,N-1} \mz_{i,N}$. These terms correspond to Wen plaquettes interacting between the $N-1$th row and the 1st row. These plaquettes however are skewed, as it is the $i-2$th and $i-1$th spin on the 1st row that interact with the $i-1$th and $i$th spin on the $N-1$th row. They all also interact with one spin on the $N$th row. With these terms, we see that we have approximately mapped a cluster state on an $N\times N$ lattice into a periodic Wen model on an $N\times (N-1) $ lattice, which interacts with the $N$th row. It is important to stress however that manipulating the generators of the stablizer group only gives us a description of the ground state because if we form a Hamiltonian by summing the new generators, there will be a different energy spectrum in the excited states. This exchange only allows us to better interpret the ground state.

Finally, one has to consider the stablizers mapped from the $ C_{i,1}$ terms. In the Wen picture, these terms all map into non-contractible loops across the $N\times (N-1)$ periodic lattice in the Wen picture. Then, locally mapping the system into the toric code picture reveals that the $ C_{i,1} $ terms map into non-contractible loops of $\mx$ operators when $i$ is odd and non-contractible loops of $ \mz$ operators when $i$ is even.

It is important that under unitary transformation the energy spectrum is conserved. In reference \cite{Dennis}, these non-contractible loops are used to store information on a topological memory by decreasing the number of available degenerate ground states. It is these loops that conserve the degeneracy of the ground state, as we are mapping from a cluster-state Hamiltonian which has a non-degenerate ground state.

\section{Conclusion}
In this work we have reviewed and proven relations between four one-and two-dimensional lattice spin models, namely the two-dimensional Cluster State model, Wen's model, Kitaev's Toric Code and the one-dimensional Ising model. These mappings can be utilized in a plethora of ways. Perhaps the most interesting one is the comparison between the two-dimensional cluster state and the Kitaev model. Both models can be used as a possible realization of quantum computation by making use of two very different ground state properties of the system, i.e. the entanglement and the toplocial structure. In this respect, we believe that our work can contribute to new insights into questions relating the fields of quantum information science and condensed matter physics. For example, can we find a simple way to characterize condensed matter systems that can be used for quantum computation?  \cite{Van}.

The other important mapping is that between the Wen model and the Ising model. Since features of topologically ordered models such as the entanglement structure can be difficult to understand intuitively, a mapping between a topologically ordered model into something simpler like the Ising model could improve our understanding of the behavior of such a system. We believe that an extension of the mapping presented in this work to lattices with arbitrary dimension will be a further step to achieve a better knowledge of the behavior of topologically ordered systems.

\noindent {\bf Acknowledgments. } We acknowledge A. Kay, S. Barrett and M.C. Ba\~{n}uls for their useful comments. This work has been supported by the National Research Foundation, Ministry of Education, Singapore, the European Community (IP- SOLID), CVK is funded by the EU project QUEVADIS and BJB is supported by the EPSRC funded Controlled Quantum Dynamics Centre for Doctoral Training.


\section*{References}
\begin{thebibliography}{10}

\bibitem{Briegel09} H. J. Briegel, D. E. Browne, W. D\"{u}r, R. Raussendorf, M. Van den Nest, Nature Physics {\bf 5}, 19 (2009).
\bibitem{Raussendorf01} R. Raussendorf and H. J. Briegel, PRL {\bf 86}, 5188 (2001)
\bibitem{Sarma05} S. D. Sarma, M. Freedman, and C. Nayak, Phys. Rev. Lett. {\bf 94}, 166802 (2005).
\bibitem{Nayak}  C. Nayak, S. H. Simon, A. Stern, M. Freedman and S. Das Sarma, Rev. Mod. Phys. {\bf 80}, 1083 (2008)
\bibitem{Kitaev03} A. Kitaev, Annals Phys. {\bf 303}, 2 (2003).
\bibitem{Dennis} E. Dennis, A. Kitaev, A. Landahl and J. Preskill, J. Math. Phys. {\bf 43}, 4452 (2002).
\bibitem{Nussinova09} Z. Nussinov, G. Ortiz, Annals of Physics {\bf 324}, 977-1057 (2009).
\bibitem{Chen07} H. Chen and J. Hu, Phys. Rev. B {\bf 76}, 193101 (2007).
\bibitem{Son} W. Son, L. Amico, R. Fazio, S. Pascazio and V. Vedral, arXiv:1103.0251v1 (2011).
\bibitem{Wenmodel} X. G. Wen, Phys. Rev. Lett. {\bf 90}, 016803 (2003).
\bibitem{Jordan28} P. Jordan and E. Wigner, Z. Physik {\bf 47}, 631 (1928)
\bibitem{Katsura61} S. Katsura, Phys. Rev {\bf 127} 1508 (1961)
\bibitem{Damian07}  D. Markham, A. Miyake and S. Virmani, New J. Phys. {\bf 9}, 194 (2007).
\bibitem{Vedral02} V. Vedral, Rev. Mod. Phys. {\bf 74}, 197 (2002).
\bibitem{Jungnitsch10} B. Jungnitsch, T. Moroder and O. G\"uhne arXiv:1010.6049 (2010).
\bibitem{Levin06} M. Levin and X. G. Wen, Phys. Rev. Lett. {\bf 96}, 110405 (2006).
\bibitem{Kitaev06} A. Kitaev and J. Preskill, Phys. Rev. Lett. {\bf 96}, 110401 (2006).
\bibitem{Plenio} M. B. Plenio J. Mod. Opt. {\bf 54}, 349 (2007)
\bibitem{Bravyi} S. Bravyi and M. B. Hastings and F. Verstaete, PRL {\bf 97}, 050401 (2006)
\bibitem{GHZ} D. M. Greenberger, M. A. Horne and A. Zeilinger, arXiv:0712.0921v1 [quant-ph] (1989)
\bibitem{NandC} M. Nielsen and I. Chuang, {\it Quantum Computation and Quantum Information}, Cambridge University Press (2000).
 \bibitem{Kitaev09} A. Kitaev, C. Laumann, arXiv:0904.2771 (2009).
\bibitem{Van} M. Van den Nest, W. DŸr, A. Miyake and H. J. Briegel, New J. Phys. {\bf 9} 204 (2007).



















\end{thebibliography}
\end{document}